# Plasmonics without losses?
# (the case for terahertz superconducting plasmonics)

*Anagnostis Tsiatmas, Vassily A. Fedotov and Nikolay I. Zheludev*

Optoelectronics Research Centre and Centre for Photonic Metamaterials, University of Southampton, Southampton SO17 1BJ, UK

*Superconductors support low-loss plasmon waves that could become information carriers in compact terahertz data processing circuits.*

The field of plasmonics, which deals with the optical properties of metallic nanostructures, is one of the most fascinating and fast-moving areas of photonics[1]. Its explosive growth in recent years has been driven by parallel advances in nano-fabrication technologies as well as a wealth of potential applications in areas ranging from bio-chemical sensing to solar power generation. Particular interest in surface plasmons –bound oscillations of electrons and light propagating at a metal surface – is based largely on the possibility that they could act as information carriers in highly-integrated nanophotonic devices transforming the chip-scale data transport paradigm by bridging the gaps between current electronic and photonic technologies. Plasmonic response is also a dominant factor underpinning functionality of photonic metamaterials[2]. However, the high promise of plasmonics is hampered by losses, an inherited feature of metal-based waveguides supporting propagating plasmon signals. Losses also create a major problem for developing photonic metamaterial technologies. This challenge brought about the intense search for solutions such as plasmon loss compensation with gain[3] and search for new, better plasmonic materials[4].

We now argue that recent developments in the field of superconducting metamaterials [5,6,7,8] and clear demonstration of phenomena such as extraordinary transmission in perforated superconducting films[9] bring about a practical proposition for superconducting plasmonics. Indeed, there are striking similarities between the electromagnetic response of metals at optical frequencies and superconductors at terahertz frequencies and below.

In metals exposed to radio and lower frequency electromagnetic waves the dynamics of free electrons is dominated by the numerous electron-electron collision events during each cycle of the driving field. Electron's mean velocity and thus the strength of electrical current induced by the wave are proportional to the instantaneous value of the field. This signifies the regime of the Ohm's law and is characterized by a large absolute value of the imaginary part of the metal's permittivity and a smaller real part (see Fig. 1a). In contrast, at higher optical frequencies electrons oscillate so rapidly that no collisions happen during one cycle: the collective dynamics of electron plasma is dominated by inertia of the carriers. Here we talk about plasmonic regime: the displacement currents become important and the real part of the metals permittivity begins to dominate. For silver such a transition from the Ohm's law electrodynamics to plasmonics occurs at frequencies above 1 THz and extends into the IR and visible (see Fig. 1a).

[1] The case for plasmonics. M. L. Bromgersma and V. M. Shalaev, Science 328, 440 (2010).
[2] The road ahead for metamaterials. N. I. Zheludev, Science 328, 582 (2010).
[3] Optical metamaterials—more bulky and less lossy. C. Soukoulis and M. Wegener, Science 330, 1633 (2010).
[4] Low-loss plasmonic metamaterials. A. Boltasseva and H. A. Atwater, Science 331, 290 (2010).
[5] Superconducting metamaterials. M. Ricci, N. Orloff and S. M. Anlage, Appl. Phys. Lett. 87, 034102 (2005).
[6] Temperature control of Fano resonances and transmission in superconducting metamaterials, V. A. Fedotov, A. Tsiatmas, J. H. Shi, R. Buckingham, P. de Groot, Y. Chen, S. Wang and N. I. Zheludev, Opt. Exp. 18, 9015 (2010).
[7] Observation of polaritons in $Bi_2Sr_2CaCu_2O_{8+x}$ single crystals. S.-O. Katterwe, H. Motzkau, A. Rydh, V. M. Krasnov. arXiv:1010.6172 (2010).
[8] Terahertz superconductor metamaterial. J. Gu, R. Singh, Z. Tian, W. Cao, Q. Xing, M. He, J. W. Zhang, J. Han, H. –T. Chen and W. Zhang, Appl. Phys. Lett. 97, 071102 (2010).
[9] Superconducting plasmonics and extraordinary transmission. A. Tsiatmas, A. R. Buckingham, V. A. Fedotov, S. Wang, Y. Chen, P. A. J. de Groot and N. I. Zheludev, Appl. Phys. Lett. 97, 111106 (2010).

While the response of plasmonic metals is characterized by dominating negative real part of permittivity at optical frequencies, in superconducting state such behaviour is found across most of the terahertz and sub-terahertz spectral domains (Fig. 1b). Indeed, the collective motion of "superconducting electrons" joined in Cooper pairs experiences no scattering and occurs freely through the lattice[10]. Their electrodynamic response, as in plasmonic metals, is determined by inertia. Moreover, imaginary part of permittivity is much lower than its real part. At higher frequencies superconductors become lossy as energy of the quanta is sufficient to break the Cooper pairs and destroy superconductivity. For example, for high-temperature superconductor yttrium-barium-copper-oxide (YBCO) the spectral domain of plasmonic-like properties extends from DC to few THz.

Unfortunately, electromagnetic fields at the surface of superconductors are almost completely expelled from the medium, making plasmon-polariton-like excitations loosely bounded to the surface, weekly localized and thus unsuitable for waveguiding application. For instance in YBCO at 1 THz the surface wave extends into vacuum for tens of wavelengths and thus differs very little from a free-space TEM-wave propagating along the surface.

Strong localization is a very desirable feature from the prospective of using superconducting plasmon-polaritons as information carriers. It can be significantly improved by increasing permittivity of the adjacent dielectric or/and at the expense of increased losses – either by using low quality superconducting films[11], or operating at frequencies close to the bandgap.

We argue that there is a much better way to unlock the potential of superconducting plasmonics. It is in using the waveguide configuration that "squeezes" plasmonic field. This can be achieved by trapping the plasmonic field laterally between two superconducting surfaces of a parallel-plate waveguide with a sub-wavelength gap of just few tens of nanometers, or using a nanometre-wide slit made in a thin superconducting film. For example, at 1 THz a TM-mode with the wavelength several times shorter than in free space (good for miniaturization) can propagate in a 50 nm wide gap without significant attenuation for tens of millimetres (Fig. 1d). In comparison, same-frequency TM-mode in a silver waveguide of the identical geometry is damped within the distance of one wavelength (Fig. 1c).

Low-loss superconducting slot waveguides is a promising solution for high-bandwidth terahertz and sub-terahertz plasmonic circuits. It offers key advantages of plasmonics such as strong lateral and longitudinal confinement necessary for high-density integration, but without dissipative losses that hamper optical plasmonic applications. As an added bonus, superconducting waveguides have weak dispersion allowing for high bandwidth and data rates. Note that the cryo-cooling requirement for superconductors is no longer a serious technological limitation as compact cryogenic devices are now widely deployed in telecommunications and sensing equipment.

However, we will be able to speak about plasmonics as a data handling and processing paradigm in the same way we speak about photonics, only when efficient techniques for active manipulation of plasmon signals are identified[12]. Here superconductors have another important advantage over metals: electromagnetic characteristics of superconductors may be readily altered by external stimuli such as magnetic field, optical illumination, surface currents or temperature. This opens a way for "active plasmonic" applications when the gap-plasmon signal can be efficiently controlled in data processing and interconnect applications.

In essence high temperature superconducting waveguides could offer promising avenue for developing compact terahertz data processing circuits.

---

[10]In the reverse analogy plasmonic metals can be considered behaving as high-frequency superconductors with the electrons in macroscopically coherent state created through the coupling with incident light.
[11]Observation of propagating plasma modes in a thin superconducting film. O. Buisson, P. Xavier, and J. Richard, Phys. Rev. Lett. 73, 3153 (1994).
[12]Active plasmonics: current status. K. F. MacDonald and N. I. Zheludev, Laser & Photon.Rev. 4, 562 (2010).

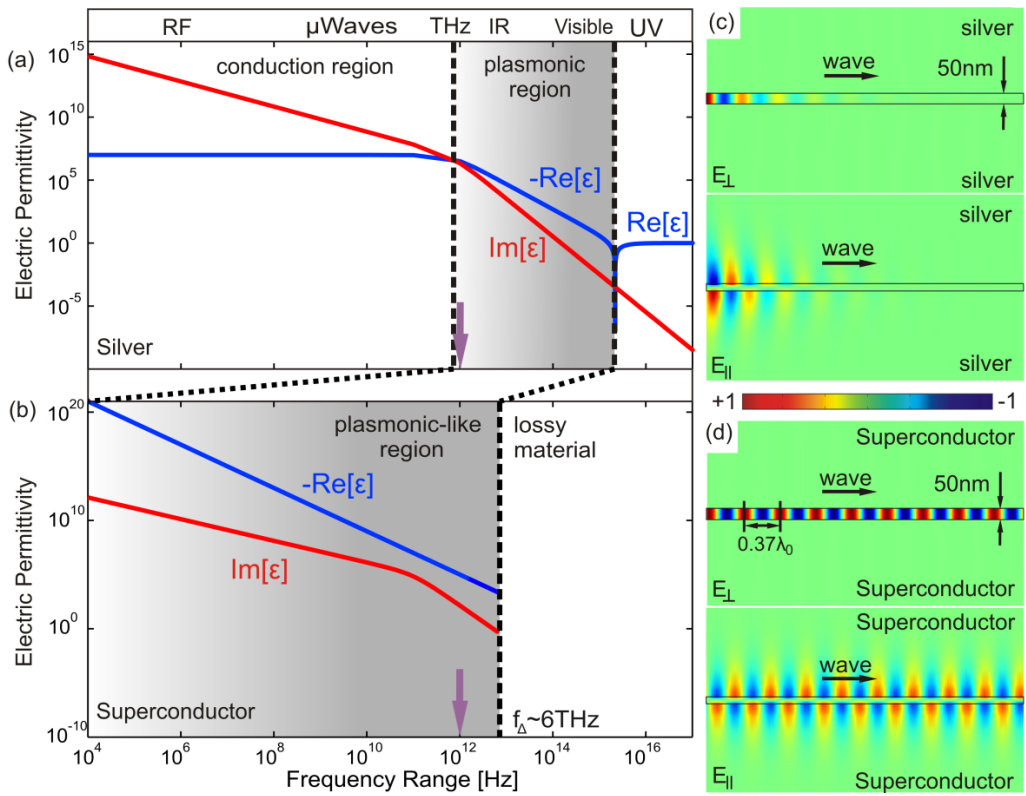

**Figure 1. Metal and superconducting plasmonic wavegudes.** Permeability of silver ε, Drude model (a), indicates plasmonic behaviour in the IR and visible parts of the spectrum. A high-temperature superconductor, two-fluid model (b), exhibits plasmonic-like behaviour at terahertz frequencies and below. Here $f_\Delta$ is the superconductor gap frequency. Distribution of electric field in a TM-wave propagating through silver (c) and superconducting (d) parallel-plate waveguides at 1 THz. $E_\parallel$ and $E_\perp$ are components of the wave's electric field parallel and perpendicular to the propagation direction. In the silver waveguide the wave decays rapidly. The wave supported by the superconducting waveguide is "compressed" (allows for compact devices) and suffers negligible losses. Here $\lambda_0 = 0.3$ mm is the free space wavelength.

e-mail: vaf@orc.soton.ac.uk; niz@orc.soton.ac.uk